\documentclass[preprint,amsmath,amssymb,aps,tightenlines]{revtex4}

\usepackage{graphicx}
\usepackage{subfigure}

\newcommand{\snuth}{{\tilde{\nu}_3}}
\newcommand{\mlight}{{\tilde{m}}}
\newcommand{\gZgZ}{g_1 Z^{1j}_N + g_2 Z^{2j}_N}

\begin{document}
\preprint{\begin{tabular}{l}
\hbox to\hsize{March, 2002 \hfill KIAS-P02007}\\[0mm]
\hbox to\hsize{hep-ph/0203251 \hfill KAIST-TH 2002/07}\\[0mm] \end{tabular} }


\bigskip

\title{Muon anomalous magnetic moment from effective supersymmetry}

\author{S. Baek}
\address{School of Physics, KIAS \\  Seoul  130-012, Korea }

\author{P. Ko and J. H. Park}
\address{Department of Physics, KAIST \\ Daejeon 305-701, Korea}

\date{March 26, 2002}

\begin{abstract}
We present a detailed analysis on the possible maximal value of the
muon $(g-2)_\mu \equiv 2 a_\mu$ within the context of effective
SUSY models with $R$ parity conservation. First of all, the mixing among
the second and the third family sleptons can contribute at one loop level
to the $a_\mu^{\rm SUSY}$ and $\tau \rightarrow \mu \gamma$ simultaneously.
One finds that the $a_\mu^{\rm SUSY}$ can be as large as
$(10 - 20 )\times 10^{-10}$ for any $\tan\beta$, imposing the upper limit on
the $\tau\rightarrow \mu \gamma$ branching ratio.
Furthermore, the two-loop Barr-Zee type contributions to $a_\mu^{\rm SUSY}$
can be significant for large $\tan\beta$, if a stop is light and $\mu$ and
$A_t$ are large enough ($\sim O(1)$ TeV).  In this case, it is possible to
have $a_\mu^{\rm SUSY}$ upto $O(10) \times 10^{-10}$
without conflicting with $\tau \rightarrow l \gamma$. We conclude
that the possible maximal value for $a_\mu^{\rm SUSY}$ is about
$\sim 20 \times 10^{-10}$ for any $\tan\beta$. Therefore the BNL experiment
on the muon $a_\mu$ can exclude the effective SUSY models only if the
measured deviation is larger than $\sim 30 \times 10^{-10}$.
\end{abstract}


\maketitle

\newpage
\narrowtext



The anomalous magnetic dipole moment (MDM) of a muon,
$a_{\mu} \equiv ( g_{\mu} - 2 )/2$, is one of the best measured quantities.
Recently, the Brookhaven E821 collaboration announced a new data on
anomalous magnetic moment $a_{\mu}$ \cite{Brown:2001mg} :
\begin{equation}
  \label{eq:amuexp}
  a_{\mu}^{\rm exp} = (11659202 \pm 14 \pm 6) \times 10^{-10}.
\end{equation}
On the other hand, the SM prediction for this quantity has been calculated
through five loops in QED and two loops in the electroweak interactions
\cite{Czarnecki:2001pv}. 
Using the corrected light--light scattering contribution to the $a_{\mu}$
through pion exchanegs \cite{kinoshita}, the difference between the data
and the SM prediction is
\begin{equation}
  \label{eq:amunew}
\delta a_\mu \equiv
a_{\mu}^{\rm exp} - a_{\mu}^{\rm SM} = (26 \pm 16) \times 10^{-10}, 
\end{equation}
which is only $1.6 \sigma$ deviation.
Therefore, the present data does not indicate any indirect evidence of new
physics at electroweak scale.
However, since the ultimate goal of the BNL experiment is to reduce the
experimental error down to $\sim 4 \times 10^{-10}$, the $\delta a_\mu$ may
provide a useful constraint on various new physics scenarios just around the
electroweak scale.

The most promising new physics beyond the SM is the minimal supersymmetric
standard model (MSSM) and its various extensions, and the muon
$(g-2)_\mu$ was one of the basic observables one considered in various
SUSY models \cite{susyamu1}.  After the BNL data was announced in the
year of 2001, there appeared a lot of works on the muon $(g-2)$ in the
context of SUSY models within the general MSSM (even with $R$
parity violation), minimal SUGRA, gauge mediation, anomaly mediation
and gaugino mediation scenarios \cite{susyamu2}.
The conclusions of these works can be summarized as follows in
a model independent manner : it is rather easy to accommodate
$\delta a_\mu \sim (10 - 70) \times 10^{-10}$ in SUSY models, if
$\mu \tan\beta$ is relatively large and SUSY particles are not too heavy.
Also the sign of the $a_\mu^{\rm SUSY}$ is correlated with the sign of the
$\mu$ parameter.

This general  conclusion seems to allude that the so-called effective
(or decoupling) SUSY models \cite{effective},
which is an attractive way to solve the SUSY
flavor and CP problems, have serious troubles if it eventually turns out
$\delta a_\mu > 10^{-10}$, since the 1st/2nd generation sfermions have to
be very heavy ($\sim O(20)$ TeV) and almost degenerate for squark sector.
One way to evade this conclusion within the effective SUSY models is simply
to invoke $R$ parity violations in order to explain the muon $(g-2)$ within
the effective SUSY models \cite{Kim:2001se}. However, the mixing between
the staus and the smuons were ignored in Ref.~\cite{Kim:2001se}, which is
not a valid assumption in generic effective SUSY models. This mixing arises
from mismatches between lepton and slepton mass matrices in the flavor space.
The effects of such a mixing among the down squarks and its effects on $B$
physics were diuscussed in Ref.~\cite{cohen} sometime ago. Our present work
is an anology of these works within the lepton sector
(see also Ref.~\cite{Cho:2001hx}). The flavor mixing
between the staus and the smuons that contribute to the $a_\mu^{\rm SUSY}$
can also enhance the decay $\tau \rightarrow \mu \gamma$, for which there
exists a new strong bound from BELLE \cite{belle} :
\[
B ( \tau \rightarrow \mu \gamma ) < 1.0 \times 10^{-6}.
\]
Thus one has to consider the $a_\mu^{\rm SUSY}$ and
$\tau \rightarrow \mu \gamma$ simultaneously.

In this letter, we present a detailed analysis on the muon $(g-2)_\mu$ in the
effective SUSY models with $R$ parity conservation, especially the possible
maximal value of $a_\mu^{\rm SUSY}$ in view of the expected new BNL data.
Lacking definite effective SUSY models, we will basically preform a numerical
analysis in a model independent way, imposing the constraint from the
unobserved decay $\tau \rightarrow \mu \gamma$. This constraint turns out
to be especially strong in the large $\tan\beta$ region. For relatively
small $\tan\beta$ (up to $\lesssim 10$), the slepton mixing allows
$a_\mu^{\rm SUSY}$ to be as large as $\sim 20\times 10^{-10}$ without having
too large $\tau \rightarrow \mu \gamma$, if there are large mixing between
the staus and smuons in both chirality sectors (namley,
$\tilde{\mu}_L - \tilde{\tau}_L$ and $\tilde{\mu}_R - \tilde{\tau}_R$
mixings). For larger $\tan\beta > 30$, the constraint from
$\tau \rightarrow \mu \gamma$ becomes very strong. Still the
$a_\mu^{\rm SUSY}$ can be as large as $9 \times 10^{-10}$ at one loop
level. Furthermore, the Barr--Zee type two loop contribution can enhance the
$a_\mu^{\rm SUSY}$ up to $(10 - 20) \times 10^{-10}$, if $A_t$ and
$\mu$ are of size $\sim O(1)$ TeV and $\tan\beta$ is large. In short, it is
not impossible to have $a_\mu^{\rm SUSY}$ as large as
$\sim 20 \times 10^{-10}$ regardless of $\tan\beta$ in effective SUSY models.
Therefore the BNL experiment on the muon $(g-2)_\mu$ can exclude the
effective SUSY models without any ambiguities only if
$\delta a_\mu > 30 \times 10^{-10}$ within the errors.


Let us first define the $l_i \rightarrow l_j \gamma$ form factors
$L_{ji}$ and $R_{ji}$ as follows :
\begin{equation}
{\cal L}_{\rm eff} (l_i \rightarrow l_j \gamma)
= e \frac{m_{l_i}}{2}  ~\overline{l}_{j} \sigma^{\mu\nu} F_{\mu\nu}
\left( L_{ji} P_L + R_{ji} P_R \right) l_i .
\end{equation}
Then, the muon $(g-2)$ or $a_\mu$ is related with $L(R)_{22}$ by
\begin{equation}
a_{\mu} = {1\over 2}~( g_{\mu} - 2 )
= m_{\mu}^2 \left( L_{22} + R_{22} \right),
\end{equation}
whereas the decay rate for $l_i \rightarrow l_{j \neq i} + \gamma$ is given by
\begin{equation}
{ Br ( l_i \rightarrow l_{j \neq i} + \gamma ) \over
Br ( l_i \rightarrow l_{j \neq i} +   \nu_i \overline{\nu}_j ) }
= {48 \pi^3 \alpha \over G_F^2 }~
\left( ~ | L_{ji} |^2 + | R_{ji} |^2 \right)
\end{equation}
We will calculate $L,R$'s relevant to $a_\mu^{\rm SUSY}$ and
$\tau \rightarrow \mu \gamma$ in the framework of effective SUSY models.
Our notations and conventions follow those of Ref.~\cite{Misiak:1997ei}.


The slepton mass matrix in the super-CKM basis is given by
\begin{equation}
  \label{eq:massmat}
  M^2_{\tilde{l}} =
  \left(
    \begin{array}{cc}
    V^E_L M^2_L V^{E\dagger}_L + m_l^2
    + \frac{\cos 2\beta}{2} ( M^2_Z - 2 M^2_W ) {\bf 1} &
    - m_l (\mu \tan\beta {\bf 1} + A_l^*)
    \\
    - m_l (\mu^* \tan\beta {\bf 1} + A_l) &
    V^E_R M^{2T}_E V^{E\dagger}_R + m_l^2
    - \cos 2\beta M^2_Z \sin^2 \theta_W {\bf 1}
    \end{array}
  \right) .
\end{equation}
This matrix is taken to be of the following form
(neglecting the trilinear couplings for charged leptons for the time being) :
\begin{equation}
    \left(
    \begin{array}{ccc|ccc}
      \tilde{m}^2_{LL11} & & & - m_e \mu \tan \beta & & \\
      & \tilde{m}^2_{LL22} & \tilde{m}^2_{LL23} & & - m_\mu
        \mu \tan \beta  & \\
      & \tilde{m}^2_{LL32} & \tilde{m}^2_{LL33} & & & -m_\tau \mu \tan \beta
     \\
      \hline
        - m_e \mu \tan \beta  & & & \tilde{m}^2_{RR11} & & \\
      & - m_\mu \mu \tan \beta  & & &
        \tilde{m}^2_{RR22} & \tilde{m}^2_{RR23} \\
      & & - m_\tau \mu \tan \beta  & &
        \tilde{m}^2_{RR32} & \tilde{m}^2_{RR33}
    \end{array}
  \right) .
\end{equation}
Since we are looking at a $CP$-conserving effect, all these
mass parameters are assumed to be real.
The origin of this kind of mixing may be the form of $M^2_{L,E}$,
the soft mass matrices
in the flavor basis, or $V^E_{L,R}$, the lepton mixing matrices.
We can diagonalize the 2-3 submatrix of the $LL$ sector into
a mixing angle $\theta_L$ and
two mass eigenvalues $\tilde{M}^2_L, \mlight^2_L$
in the limit of no $LR$ mixing:
\begin{equation}
  \left(
    \begin{array}{cc}
      \tilde{m}^2_{LL22} & \tilde{m}^2_{LL23} \\
      \tilde{m}^2_{LL32} & \tilde{m}^2_{LL33}
    \end{array}
  \right)
  =
  \left( \begin{array}{cc}
      \cos\theta_L  &  \sin\theta_L\\
      -\sin\theta_L &
      \cos\theta_L  \end{array}
  \right)
  \left(
    \begin{array}{cc}
      \tilde{M}^2_L & \\
      & \mlight^2_L
    \end{array}
  \right)
  \left( \begin{array}{cc}
      \cos\theta_L  &  -\sin\theta_L\\
      \sin\theta_L &
      \cos\theta_L  \end{array}
  \right) ,
\end{equation}
and likewise for the $RR$ sector.
The sneutrino mass matrix with the neutrino masses neglected is
\begin{equation}
  M^2_{\tilde{\nu}} = V^\nu_L M^2_L V^{\nu\dagger}_L +
  \frac{\cos 2 \beta}{2} M^2_Z {\bf 1} ,
\end{equation}
and the lightest sneutrino mass is
\begin{equation}
  m_\snuth^2 = \mlight^2_L + \cos 2 \beta M^2_W,
\end{equation}
when we also ignore the lepton masses.
If $V^\nu_L$ is different from $V^E_L$, $M^2_{\tilde{\nu}}$ is
diagonalized by a different unitary matrix than
the $LL$ sector of $M^2_{\tilde{l}}$.
However, this misalignment is compensated by the MNS matrix at the
chargino-lepton-sneutrino vertex, and the chargino amplitudes can be
expressed in terms of the slepton mixing angles, $\theta_L$ and
$\theta_R$, if we ignore neutrino and lepton masses in
$M^2_{\tilde{\nu}}$ and $M^2_{\tilde{l}}$.

The question about the sizes of $\tilde{m}_{AA33}^2$ and
$\tilde{m}_{AA23}^2$ (with $A = L, R$) is highly model dependent one,
depending on the details of the underlying model and may be closely related
with understanding flavor structures in the MSSM. 
Note that the SUSY flavor problem is stated in the super-CKM basis
as follows : the sfermion mass matrices should be flavor diagonal in
this basis and/or the sfermion masses should be almost degenerate.
Most effective SUSY models in the literature have hierarchical sfermion
mass structures (which are almost diagonal with small mixing angles among
different generations) in the flavor basis, namely $M_L^2 $ and $M_E^2$
\cite{effective}. However, it'd not be impossible to construct a
model of large flavor mixings in the second and third generation sfermions,
especially considering the large mixings in the neutrino sector.
In the super-CKM basis, the slepton mass matrices 
$M_L^2 $ and $M_E^2$ are multiplied by $V_L^E$, $V_R^E$
and $V_L^{\nu}$ with $V_{MNS} \equiv V_L^E V_L^{\nu \dagger}$. Because of
the large mixings among three light neutrinos, the resulting slepton mass
matrices can have large and comparable elements. (A similar argument may be
true for the righthanded slepton sector as well.) This is  a source
of the large mixings among the sleptons, which can enhance the
$a_\mu^{\rm SUSY}$ in the effective SUSY models.

The heavier mass eigenstates decoupling,
it is straightforward to show that the
$a_\mu^{\rm SUSY} = a_{\mu}^C + a_{\mu}^N$
is given by
\begin{eqnarray}
a_{\mu}^C & = &
\frac{2}{(4\pi)^2}
\frac{m_\mu^2}{m_\snuth^2}
\sum_j \left[
  g_2^2 |Z^+_{1j}|^2 f_1(x_j)
  - \frac{m_{C_j}}{v\cos\beta} g_2 Z^-_{2j} Z^+_{1j} f_2(x_j)
\right] \sin^2\theta_L ,
\nonumber   \\
\label{eq:amususy}
a_{\mu}^N & = &
\frac{2 m_\mu^2}{(4\pi)^2}
\sum_j \Bigg[ \nonumber \\
&& \left(
  \frac{1}{\sqrt{2}} (\gZgZ) \frac{m_{N_j}}{v\cos\beta}
  Z^{3j}_N f_4(x_{jL})
  - \frac{1}{2} |\gZgZ|^2 f_3(x_{jL})
\right) \frac{\sin^2\theta_L}{\mlight^2_L} \nonumber \\
&& - \left(
  \sqrt{2} g_1 \frac{m_{N_j}}{v\cos\beta} Z^{1j}_N Z^{3j}_N f_4(x_{jR})
  + 2 g_1^2 |Z^{1j}_N|^2 f_3(x_{jR})
\right) \frac{\sin^2\theta_R}{\mlight^2_R} \nonumber \\
&& - g_1 Z^{1j}_N (\gZgZ)
\frac{m_{N_j}\mu\tan\beta}{\mlight^2_L - \mlight^2_R}
  \left(\frac{f_4(x_{jL})}{\mlight^2_L} -
    \frac{f_4(x_{jR})}{\mlight^2_R}\right)
\frac{m_\tau}{m_\mu}
\frac{\sin2\theta_L \sin2\theta_R}{4}
\Bigg] ,
\end{eqnarray}
where $x_j \equiv m_{C_j}^2 / m_\snuth^2$,
$x_{jL(R)} \equiv m_{N_j}^2 / \mlight^2_{L(R)}$,
and $v^2 = 2 m_Z^2 / (g_1^2 + g_2^2)$.
The loop functions are defined as follows:
\begin{eqnarray}
  f_1(x) &=& \frac{1}{12 (x-1)^4}
  (2 + 3x - 6x^2 + x^3 + 6x \log x) ,
  \\
  f_2(x) &=& \frac{1}{2 (x-1)^3}
  (3 - 4x + x^2 + 2 \log x) ,
  \\
  f_3(x) &=& \frac{1}{12 (x-1)^4}
  (1 - 6x + 3x^2 + 2x^3 - 6x^2 \log x) ,
  \\
  f_4(x) &=& \frac{1}{2 (x-1)^3}
  (-1 + x^2 - 2x \log x) .
\end{eqnarray}
In the limit of no slepton flavor mixing ($\theta_L = \theta_R = \pi/2$),
we have checked that our results reduce to the previous results in the MSSM.
Let us note that the neutralino-stau loop contribution to the $a_\mu$ can be
enhanced by $m_\tau / m_\mu$ if both $\tilde{\mu}_L - \tilde{\tau}_L$ and
$\tilde{\mu}_R - \tilde{\tau}_R$ mixing are (near) maximal. On the other hand,
if the mixing is significant only in one chirality sector (namely, if
$\theta_L = 0 $ or $\theta_R = 0$), there is no such an enhancement factor,
and the resulting $a_\mu^{\rm SUSY}$ will be less than the case $\theta_L =
\theta_R = \pi/4$. This was also noticed in Ref.~\cite{Cho:2001hx}.

One can also calculate the amplitude for the decay
$\tau \rightarrow \mu \gamma$.
The coefficients relevant to this process read as
\begin{eqnarray}
  L^C_{23} & = &
  \frac{1}{(4\pi)^2}
  \frac{1}{m_\snuth^2}
  \sum_j \left[
    g_2^2 |Z^+_{1j}|^2 f_1(x_j)
    - \frac{m_{C_j}}{v\cos\beta} g_2 Z^-_{2j} Z^+_{1j} f_2(x_j)
  \right] \frac{m_\mu}{m_\tau} \frac{\sin2\theta_L}{2} ,
  \\
  L^N_{23} & = &
  \frac{1}{(4\pi)^2}
  \sum_j \Bigg[ \nonumber \\
  && \left(
    \frac{1}{\sqrt{2}} (\gZgZ) \frac{m_{N_j}}{v\cos\beta}
    Z^{3j}_N f_4(x_{jL})
    - \frac{1}{2} |\gZgZ|^2 f_3(x_{jL})
  \right) \frac{m_\mu}{m_\tau} \frac{\sin2\theta_L}{2 \mlight^2_L} \nonumber \\
  && - \left(
    \sqrt{2} g_1 \frac{m_{N_j}}{v\cos\beta} Z^{1j}_N Z^{3j}_N
    f_4(x_{jR})
    + 2 g_1^2 |Z^{1j}_N|^2 f_3(x_{jR})
  \right) \frac{\sin2\theta_R}{2 \mlight^2_R} \nonumber \\
  && - g_1 Z^{1j}_N (\gZgZ)
  \frac{m_{N_j}\mu\tan\beta}{\mlight^2_L - \mlight^2_R}
  \left(\frac{f_4(x_{jL})}{\mlight^2_L} -
    \frac{f_4(x_{jR})}{\mlight^2_R}\right)
  \frac{\cos^2\theta_L \sin2\theta_R}{2}
  \Bigg]
\end{eqnarray}
\begin{eqnarray}
  R^{C*}_{23} & = &
  \frac{1}{(4\pi)^2}
  \frac{1}{m_\snuth^2}
  \sum_j \left[
    g_2^2 |Z^+_{1j}|^2 f_1(x_j)
    - \frac{m_{C_j}}{v\cos\beta} g_2 Z^-_{2j} Z^+_{1j} f_2(x_j)
  \right] \frac{\sin2\theta_L}{2} ,
  \\
  \label{eq:r23susy}
  R^{N*}_{23} & = &
  \frac{1}{(4\pi)^2}
  \sum_j \Bigg[ \nonumber \\
  && \left(
    \frac{1}{\sqrt{2}} (\gZgZ) \frac{m_{N_j}}{v\cos\beta}
    Z^{3j}_N f_4(x_{jL})
    - \frac{1}{2} |\gZgZ|^2 f_3(x_{jL})
  \right) \frac{\sin2\theta_L}{2 \mlight^2_L} \nonumber \\
  && - \left(
    \sqrt{2} g_1 \frac{m_{N_j}}{v\cos\beta} Z^{1j}_N Z^{3j}_N
    f_4(x_{jR})
    + 2 g_1^2 |Z^{1j}_N|^2 f_3(x_{jR})
  \right) \frac{m_\mu}{m_\tau} \frac{\sin2\theta_R}{2 \mlight^2_R} \nonumber \\
  && - g_1 Z^{1j}_N (\gZgZ)
  \frac{m_{N_j}\mu\tan\beta}{\mlight^2_L - \mlight^2_R}
  \left(\frac{f_4(x_{jL})}{\mlight^2_L} -
    \frac{f_4(x_{jR})}{\mlight^2_R}\right)
  \frac{\cos^2\theta_R \sin2\theta_L}{2}
  \Bigg]
\end{eqnarray}

In order that our numerical analysis be as model independent as possible,
we fixed $\tilde{m}_{LL22} = \tilde{m}_{RR22} = 10$ TeV,
and scanned the following parameter range :
\begin{align}
2 \le \tan \beta \le 50,
~~~ 0.2 \text{ TeV} \le \mu&, ~M_2 \le 1 \text{ TeV}~~~, \nonumber\\
(0.1 \text{ TeV})^2 \le \tilde{m}^2_{LL33},~~\tilde{m}^2_{RR33}
&\le ( 10 \text{ TeV} )^2, \label{eq:range} \\
- ( 10 \text{ TeV} )^2 \le \tilde{m}^2_{LL23},~~
\tilde{m}^2_{RR23} &\le + ( 10 \text{ TeV} )^2.  \nonumber
\end{align}
Note that the effective SUSY models do not necessarily imply that the slepton
mass parameters $ \tilde{m}^2_{LL33}$ and/or $ \tilde{m}^2_{RR33}$ should be
(electroweak scale)$^2$. Since slepton Yukawa couplings are small, their
effects on the one loop corrected Higgs mass are  negligible. Therefore staus
and tau sneutrinos need not be light in the effective SUSY models.
However the resulting $a_\mu^{\rm SUSY}$ will be very small for very heavy
staus and tau sneutrinos.
Then we selected parameter sets yielding positive slepton (mass)$^2$ and
satisfying the direct search bounds : $m_{\tilde{\tau}} \ge 85$ GeV, 
$m_{\tilde{\nu}_3} > 44.7$ GeV and $m_{\chi^+} > 103.5$ GeV 
\cite{sparticles:limit}. We used the GUT relation $M_1 / M_2 = 5 \alpha_1
/ 3 \alpha_2$ to fix $M_1$ for a given $M_2$.  The trilinear couplings for
the charged leptons are set to zero. For large $\tan\beta$, the trilinear
couplings are almost irrelevant. For small and moderate $\tan\beta$, it
changes the $LR$ mixing parameters, and we have checked the
constrained maximal $a_\mu^{\rm SUSY}$
can change upto $\pm 2 \times 10^{-10}$ when we varied the $A_l$'s from
$-1$ TeV to $+1$ TeV.

In Fig.~1, we show the possible maximal value of $a_\mu^{\rm SUSY}$ (at one
loop level) as a function of $\tan\beta$ with and without the
$\tau\rightarrow \mu \gamma$ constraint in solid and dotted curves,
respectively. If $\tan\beta$ is not too large, the
$\tau\rightarrow \mu \gamma$ constraint does not overkill the
$a_\mu^{\rm SUSY}$. For large $\tan\beta$, the one loop contribution to
$a_\mu^{\rm SUSY}$ can be much larger, but is strongly constrained by
$\tau \rightarrow \mu \gamma$. Still the resulting $a_\mu^{\rm SUSY}$ can be
as large as $9 \times 10^{-10}$.
This point is also illustrated by Figs.~\ref{fig2} (a) and (b), where we
show the region plots for $\tan\beta = 3$ (a) and $\tan\beta = 30$ (b).
In Fig.~\ref{fig2} (a), $a_\mu^{\rm SUSY}$ can reach $O(20\times10^{-10})$
for $\tan\beta = 3$, still satisfying the $\tau\rightarrow \mu \gamma$
constraint.  For $\tan\beta = 30$, we have
$a_\mu^{\rm SUSY} \lesssim 10\times10^{-10}$ [Fig.~2 (b)].
This behavior can be easily understood, since $a_\mu^{\rm SUSY} \propto
\tan\beta$ whereas $B ( \tau\rightarrow \mu \gamma ) \propto \tan^2 \beta$.
Therefore the constraint becomes much more severe when $\tan\beta $ is large,
in which the one loop $a_\mu^{\rm SUSY}$ is essentially smaller than
$10 \times 10^{-10}$. 
The possible maximal value for $a_\mu^{\rm SUSY}$ will decrease as the upper 
limit on $B ( \tau\rightarrow \mu \gamma )$ gets improved.  

In the effective SUSY models, the two-loop contributions to the EDM's and
MDM's through a third (s)fermion loop could be substantial for large
$\tan\beta$ \cite{Chen:2001kn,Arhrib:2001xx}.
Since previous discussion implies that the one loop contribution to
$a_\mu^{\rm SUSY}$ cannot be larger than $\sim 10 \times 10^{-10}$ for large
$\tan\beta$ in the effective SUSY models, it is important to estimate these
two loop contributions which may dominate in the large $\tan\beta$ region.
The basic formulae for these contributions have been derived both for the
neutral and the charged Higgs exchanges with (s)top and/or (s)bottom loops :
\begin{equation}
a_{\mu}^{\rm 2-loop} = - {\alpha \over 2 \pi}~\left( { G_F m_{\mu}^2 \over
4 \sqrt{2} \pi^2 } \right)~\lambda_{\mu}^S \sum_{\tilde{f}}~N_c^{\tilde{f}}
Q_{\tilde{f}}^2 ~{\lambda_{\tilde{f}} \over m_S^2}~{\cal F} ( m_{\tilde{f}}^2
/ m_S^2 )
\end{equation}
where $N_c^{\tilde{f}}$, $Q_{\tilde{f}}$ and $m_{\tilde{f}}$ are the number
of colors, the electric charge and the mass of the internal sfermion in the
loop, and $m_S$ (with $S = h^0$ or $H^0$) is the mass of the exchanged scalar
Higgs $h^0$ or $H^0$. $\lambda_\mu^{( h^0 , H^0 )} = ( -\sin\alpha ,
\cos \alpha ) / \cos\beta$, where $\alpha$ is the mixing angle of neutral
CP-even Higgs bosons. The explicit form of the loop function ${\cal F}(z)$
can be found in Ref.~\cite{Arhrib:2001xx}.
Note that the expression in the parenthesis,
\[
{G_F m_\mu^2 \over 4 \sqrt{2} \pi^2} = 23.3 \times 10^{-10}
\]
is the size of the SM electroweak corrections to the muon $(g-2)_\mu$, and thus
the above two-loop Barr-Zee type contributions to the muon $(g-2)_\mu$ can be
substantial for large $\tan\beta$, and the large positive $\mu$ or the large
negative $A_f$. The larger the trilinear coupling $A_t$ is, the larger
$(g-2)_\mu$ one can afford.

In Fig.~1, we also show the two-loop Barr-Zee type contribution to
$a_\mu^{\rm SUSY}$ for three different $\mu = 0.5, 1 $ and 2 TeV's
(the long dashed, the dot-dashed and the short dashed curves, respectively).
We have assumed the maximal mixing angle for neutral Higgs bosons, and set
$m_S = 100$ GeV (just above the current lower limit on the CP-even heavier 
neutral Higgs boson $H$) in order to maximize the desired effect.
There is a clear evidence that this two-loop effects becomes
important as $\tan\beta$ grows. Adding the two-loop Barr-Zee type contribution
to the one loop effects, the possible maximal value for $a_\mu^{\rm SUSY}$
can easily extend to $(20 - 30) \times 10^{-10}$ even for large $\tan\beta$.
Therefore it'd not be possible to completely rule out the effective SUSY
models from the BNL experiment on the muon MDM, unless the deviation between
the SM prediction and the data is larger than, say, $\sim 30 \times 10^{-10}$.

We also plot the dependence of the possible maximal value of
$a_\mu^{\rm SUSY}$ on the SUSY breaking parameter
$\tilde{m}_{LL33} = \tilde{m}_{RR33} = \tilde{m}_{33}$ in Fig.~3 for
$\tan\beta = 3, 10$ and 40, respectively. The lower (the upper) curves are
with (without) $\tau \rightarrow \mu \gamma$ constraint.
A larger value of $a_\mu^{\rm SUSY}$ is possible, if $\tilde{m}_{33}$
becomes larger. The reason lies in that in this case one needs a large mixings
$\tilde{m}_{LL23}^2$ and $\tilde{m}_{RR23}^2$ in order to have light stops
at the electroweak scale if $\tilde{m}_{33}$ becomes large.
(Note that we had fixed  $\tilde{m}_{LL22} = \tilde{m}_{RR22} = 10$ TeV
and we need light stops
around a few hundred GeV's in order to have a significant effect on the muon
$(g-2)$.) Therefore the $a_\mu^{\rm SUSY}$ in the effective SUSY models can be
$\sim 20 \times 10^{-10}$ at one loop level,
if $\tan\beta$ is not too large and the slepton
mass parameters involving the 3rd generations are also very large (upto
$O({\rm few} - 10)$ TeV) so that one can have light slepton spectra and large
mixings.
On the other hand, if one naively apply the idea of light staus directly to
the mass parameters $\tilde{m}_{LL33}^2$ and $\tilde{m}_{RR33}^2$ (and
necessarily with small flavor mixings $\tilde{m}_{LL23}^2$ and
$\tilde{m}_{RR23}^2$ in order to have light but non-tachyonic stops),
the resulting $a_\mu^{\rm SUSY}$ cannot be large :
$a_{\mu}^{\rm SUSY} \lesssim 3 \times 10^{-10}$ if $\tilde{m}_{LL33} =
\tilde{m}_{RR33} < O(1)$ TeV, for example (see Fig.~3).

\begin{figure}
\centering
\includegraphics[width=12cm]{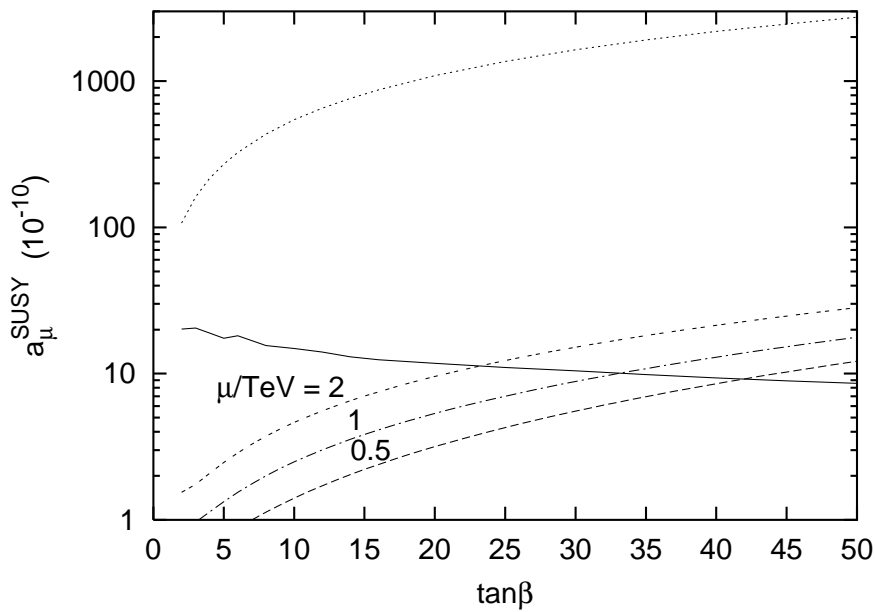}
\caption{The possible maximal value of $a_\mu^{\rm SUSY}$ at one loop
order in the effective SUSY models as a function of $\tan\beta$,
with and without the $\tau\rightarrow \mu \gamma$ constraint
(the solid and the dotted curves, respectively). The lower three curves
represent the two-loop Barr-Zee type contributions to $a_\mu^{\rm SUSY}$ for
$m_S = 100$ GeV and the maximal mixing angle for neutral Higgs bosons. }
\label{fig1}
\end{figure}

\begin{figure}
\centering
\subfigure[$\tan \beta = 3$]{%
\includegraphics[height=7cm]{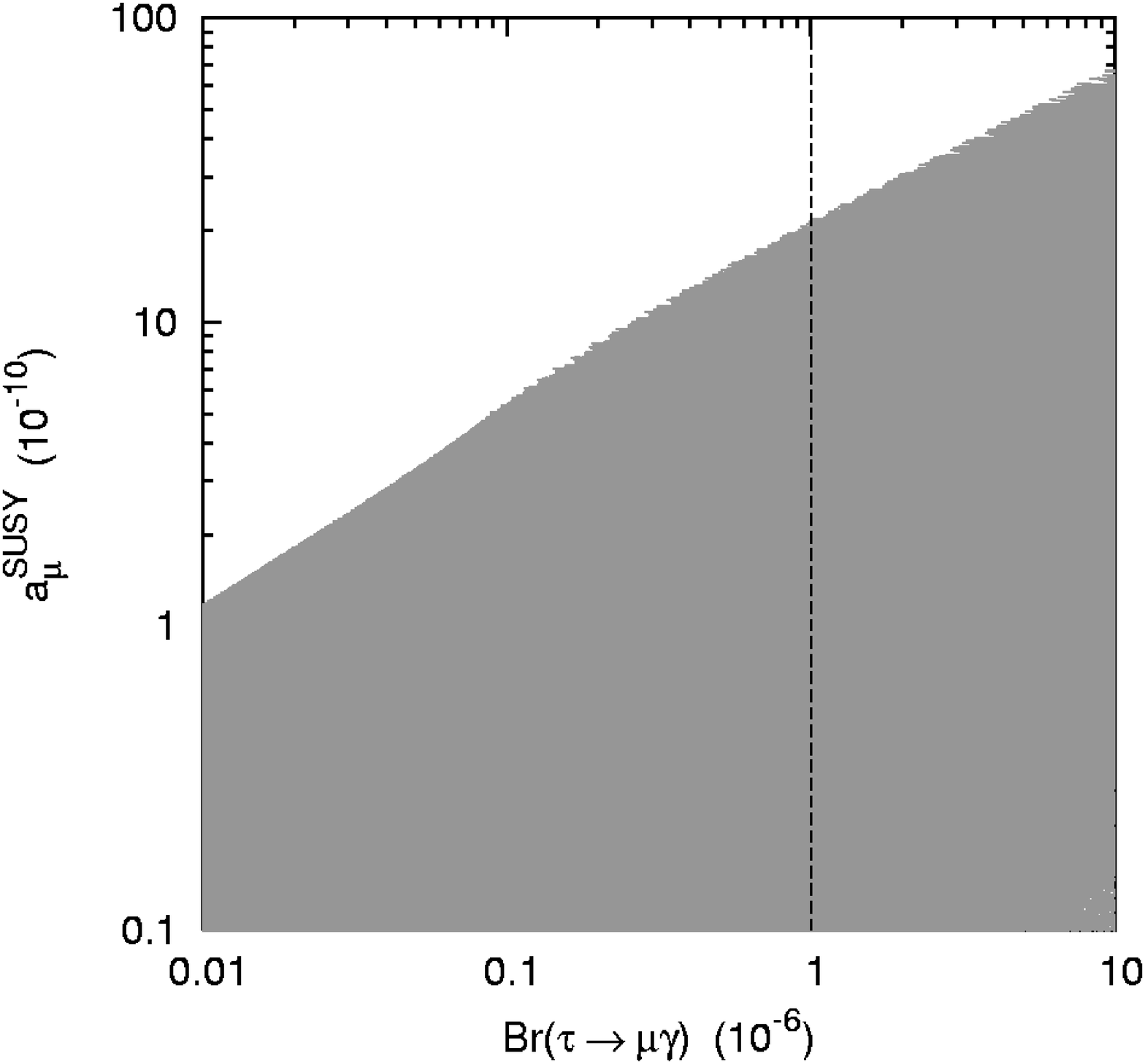}}\hspace{5mm}
\subfigure[$\tan \beta = 30$]{%
\includegraphics[height=7cm]{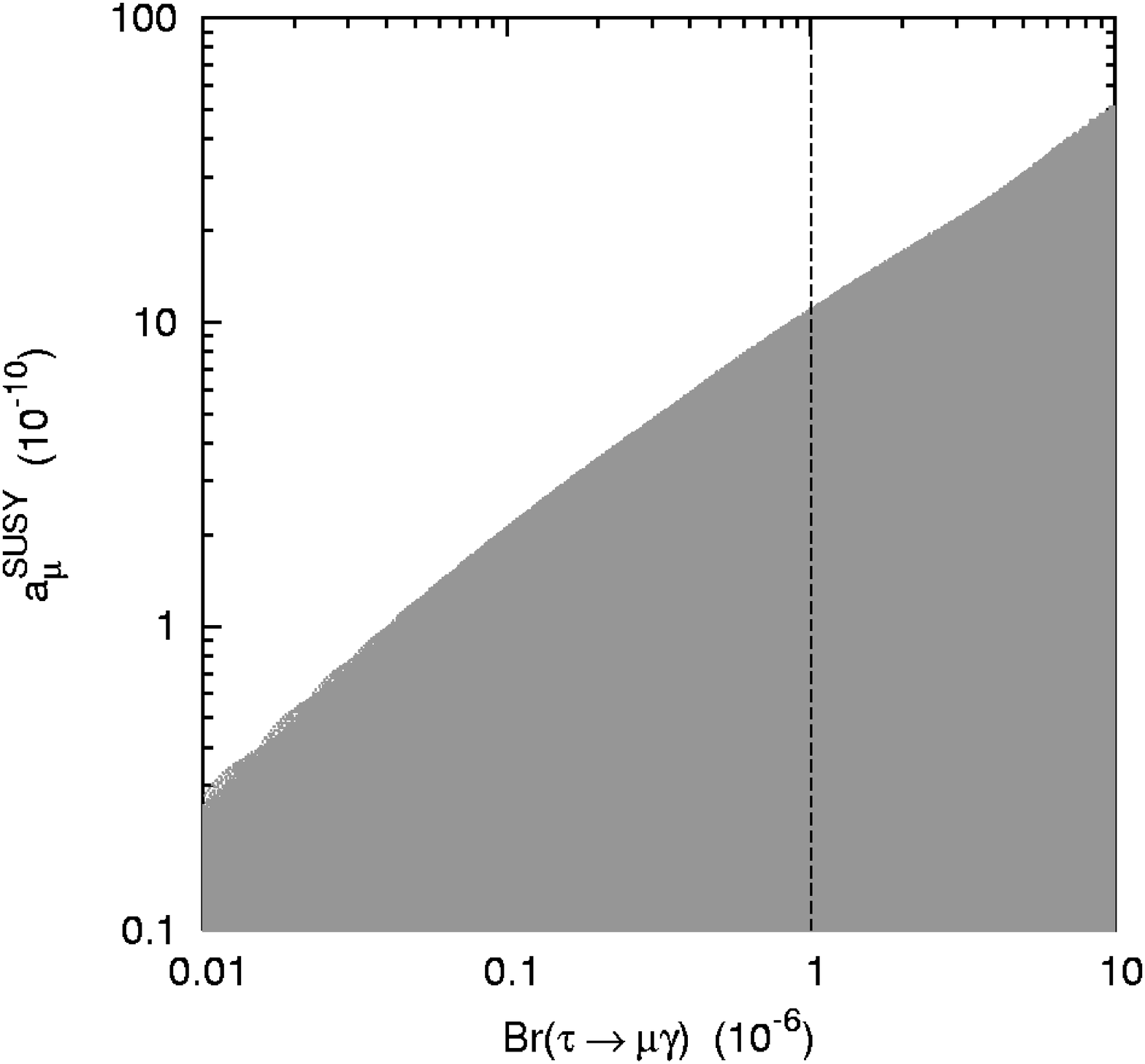}}
\caption{Regions on $a_\mu^\mathrm{SUSY}$--$Br( \tau \rightarrow \mu \gamma )$
plane swept as the parameters are varied within Range (\ref{eq:range})
with $\tan\beta$ fixed at 3 and 30.
The vertical dashed line shows upper bound on the branching ratio of
90\% confidence level.
}
\label{fig2}
\end{figure}

\begin{figure}
\centering
\includegraphics[width=12cm]{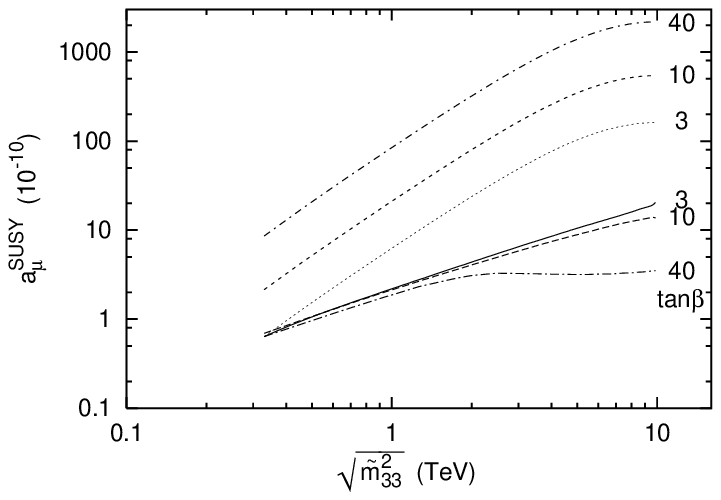}
\caption{
The possible maximal value of $a_\mu^{\rm SUSY}$ as a function of
$\tilde{m}_{33} = \tilde{m}_{LL33} = \tilde{m}_{RR33}$,
with and without the $\tau\rightarrow \mu \gamma$
constraint (the lower and the upper curves, respectively).
}
\label{fig3}
\end{figure}

Note that the maximum of $\tan\beta = 40$ curve in Fig.~\ref{fig3} is lower
than the $\tan\beta = 40$ point of Fig.~\ref{fig1} about $10\times10^{-10}$.
This is because $\tilde{m}_{LL33}$ and $\tilde{m}_{RR33}$ were assumed to be
equal in Fig.~\ref{fig3}, but not in Fig.~\ref{fig1}.
It turns out that in small $\tan\beta$ case, $a_{\mu}^{\rm SUSY}$
gets maximized when $\mlight^2_L \simeq \mlight^2_R$ and
$\theta_L \simeq \theta_R \simeq \pi/4$, while in large $\tan\beta$ case,
$\mlight^2_L / \mlight^2_R \simeq 60$ and
$\theta_L \simeq 0.18, \theta_R \simeq \pi/4$.
Let us note another point here.
The result that $a_{\mu}^{\rm SUSY}$ reaches $9\times10^{-10}$
when $\tan\beta = 40$, was obtained from
Eqs.~(\ref{eq:amususy}--\ref{eq:r23susy}).
If we treat the $LR$ mixing by fully diagonalizing the $4\times4$
mass matrix, this maximal number gets reduced to $6\times10^{-10}$.



In conclusion, we considered the muon $(g-2)_\mu$ within the effective SUSY
models. In this case, the smuon and the muon sneutrino loop contributions
to the muon $(g-2)_\mu$ is negligible. However, the staus can contribute
to the muon $(g-2)_\mu$ through the flavor mixing in the slepton sector.
Including the current constraint from $\tau \rightarrow \mu \gamma$, we
find that $a_\mu^{\rm SUSY}$ in the effective SUSY model can be as large as
$\sim 20 \times 10^{-10}$ in a reasonable region of parameter space.
This bound is fairly model independent witin the effective SUSY models, and
will become smaller once the upper bounds on $\tau \rightarrow \mu \gamma$
is improved.
Our study shows that the $a_\mu^{\rm SUSY}$ can be as large as $\sim 20 \times
10^{-10}$ in the effective SUSY models for all $\tan\beta$ if there is a
large mixing between the second and third generation sfermions. For large
$\tan\beta$, the constraint from $\tau\rightarrow \mu\gamma$ is very strong
but $a_\mu^{\rm SUSY}$ can be as large as $9 \times 10^{-10}$. Also it can
receive additional contributions from two-loop Barr-Zee type contributions
of the similar size. Overall, the possible maximal value for
$a_\mu^{\rm SUSY}$ is about $20 \times 10^{-10}$ so that the BNL experiment
on the muon $(g-2)_\mu$ can exclude the effective SUSY models only if the
measured deviation is larger than $\sim 30 \times 10^{-10}$.

\acknowledgements
We are grateful to Kiwoon Choi and Wan Young Song for useful discussions.
This work is supported in part by BK21 Core program of the Ministry of
Education (MOE), and by the Korea Science and Engineering Foundation (KOSEF)
through Center for High Energy Physics (CHEP) at Kyungpook
National University.


\begin{thebibliography}{99}
\bibitem{Brown:2001mg}
H.~N.~Brown {\it et al.}  [Muon g-2 Collaboration],
Phys.\ Rev.\ Lett.\  {\bf 86}, 2227 (2001)
[arXiv:hep-ex/0102017].
\bibitem{Czarnecki:2001pv}
A.~Czarnecki and W.~J.~Marciano,
Phys.\ Rev.\ D {\bf 64}, 013014 (2001)
[arXiv:hep-ph/0102122].
\bibitem{kinoshita}
M.~Knecht and A.~Nyffeler,
arXiv:hep-ph/0111058 ;
M.~Knecht, A.~Nyffeler, M.~Perrottet and E.~De Rafael,
arXiv:hep-ph/0111059 ;
M.~Hayakawa and T.~Kinoshita,
arXiv:hep-ph/0112102 ;
J.~Bijnens, E.~Pallante and J.~Prades,
arXiv:hep-ph/0112255.
\bibitem{susyamu1}
J.~R.~Ellis, J.~S.~Hagelin and D.~V.~Nanopoulos,
Phys.\ Lett.\ B {\bf 116}, 283 (1982) ;
R.~Barbieri and L.~Maiani,
Phys.\ Lett.\ B {\bf 117}, 203 (1982) ;
D.~A.~Kosower, L.~M.~Krauss and N.~Sakai,
Phys.\ Lett.\ B {\bf 133}, 305 (1983) ;
J.~L.~Lopez, D.~V.~Nanopoulos and X.~Wang,
Phys.\ Rev.\ D {\bf 49}, 366 (1994)
[arXiv:hep-ph/9308336] ;
U.~Chattopadhyay and P.~Nath,
Phys.\ Rev.\ D {\bf 53}, 1648 (1996)
[arXiv:hep-ph/9507386] ;
T.~Moroi,
Phys.\ Rev.\ D {\bf 53}, 6565 (1996)
[Erratum-ibid.\ D {\bf 56}, 4424 (1996)]
[arXiv:hep-ph/9512396] ;
M.~Carena, G.~F.~Giudice and C.~E.~Wagner,
Phys.\ Lett.\ B {\bf 390}, 234 (1997)
[arXiv:hep-ph/9610233] ;
K.~T.~Mahanthappa and S.~Oh,
Phys.\ Rev.\ D {\bf 62}, 015012 (2000)
[arXiv:hep-ph/9908531].
\bibitem{susyamu2}
L.~L.~Everett, G.~L.~Kane, S.~Rigolin and L.~T.~Wang,
Phys.\ Rev.\ Lett.\  {\bf 86}, 3484 (2001)
[arXiv:hep-ph/0102145] ;
J.~L.~Feng and K.~T.~Matchev,
Phys.\ Rev.\ Lett.\  {\bf 86}, 3480 (2001)
[arXiv:hep-ph/0102146] ;
E.~A.~Baltz and P.~Gondolo,
Phys.\ Rev.\ Lett.\  {\bf 86}, 5004 (2001)
[arXiv:hep-ph/0102147] ;
U.~Chattopadhyay and P.~Nath,
Phys.\ Rev.\ Lett.\  {\bf 86}, 5854 (2001)
[arXiv:hep-ph/0102157] ;
R.~Arnowitt, B.~Dutta, B.~Hu and Y.~Santoso,
Phys.\ Lett.\ B {\bf 505}, 177 (2001)
[arXiv:hep-ph/0102344] ;
S.~Komine, T.~Moroi and M.~Yamaguchi,
Phys.\ Lett.\ B {\bf 506}, 93 (2001)
[arXiv:hep-ph/0102204] ;
J.~R.~Ellis, D.~V.~Nanopoulos and K.~A.~Olive,
Phys.\ Lett.\ B {\bf 508}, 65 (2001)
[arXiv:hep-ph/0102331] ;
J.~Hisano and K.~Tobe,
Phys.\ Lett.\ B {\bf 510}, 197 (2001)
[arXiv:hep-ph/0102315] ;
K.~Choi, K.~Hwang, S.~K.~Kang, K.~Y.~Lee and W.~Y.~Song,
Phys.\ Rev.\ D {\bf 64}, 055001 (2001)
[arXiv:hep-ph/0103048] ;
S.~P.~Martin and J.~D.~Wells,
Phys.\ Rev.\ D {\bf 64}, 035003 (2001)
[arXiv:hep-ph/0103067] ;
S.~Komine, T.~Moroi and M.~Yamaguchi,
Phys.\ Lett.\ B {\bf 507}, 224 (2001)
[arXiv:hep-ph/0103182] ;
S.~w.~Baek, P.~Ko and H.~S.~Lee,
Phys.\ Rev.\ D {\bf 65}, 035004 (2002)
[arXiv:hep-ph/0103218] ;
D.~F.~Carvalho, J.~R.~Ellis, M.~E.~Gomez and S.~Lola,
Phys.\ Lett.\ B {\bf 515}, 323 (2001)
[arXiv:hep-ph/0103256] ;
H.~Baer, C.~Balazs, J.~Ferrandis and X.~Tata,
Phys.\ Rev.\ D {\bf 64}, 035004 (2001)
[arXiv:hep-ph/0103280] ;
S.~w.~Baek, T.~Goto, Y.~Okada and K.~i.~Okumura,
Phys.\ Rev.\ D {\bf 64}, 095001 (2001)
[arXiv:hep-ph/0104146] ;
G.~C.~Cho and K.~Hagiwara,
Phys.\ Lett.\ B {\bf 514}, 123 (2001)
[arXiv:hep-ph/0105037].
\bibitem{effective}
M.~Dine, A.~Kagan and S.~Samuel,
Phys.\ Lett.\ B {\bf 243}, 250 (1990) ;
S.~Dimopoulos and G.~F.~Giudice,
Phys.\ Lett.\ B {\bf 357}, 573 (1995)
[arXiv:hep-ph/9507282] ;
A.~Pomarol and D.~Tommasini,
Nucl.\ Phys.\ B {\bf 466}, 3 (1996)
[arXiv:hep-ph/9507462] ;
A.~G.~Cohen, D.~B.~Kaplan and A.~E.~Nelson,
Phys.\ Lett.\ B {\bf 388}, 588 (1996)
[arXiv:hep-ph/9607394] ;
D.~E.~Kaplan, F.~Lepeintre, A.~Masiero, A.~E.~Nelson and A.~Riotto,
Phys.\ Rev.\ D {\bf 60}, 055003 (1999)
[arXiv:hep-ph/9806430] ;
J.~Hisano, K.~Kurosawa and Y.~Nomura,
Phys.\ Lett.\ B {\bf 445}, 316 (1999)
[arXiv:hep-ph/9810411] ;
J.~Hisano, K.~Kurosawa and Y.~Nomura,
Nucl.\ Phys.\ B {\bf 584}, 3 (2000)
[arXiv:hep-ph/0002286].
\bibitem{Kim:2001se}
J.~E.~Kim, B.~Kyae and H.~M.~Lee,
Phys.\ Lett.\ B {\bf 520}, 298 (2001)
[arXiv:hep-ph/0103054].
\bibitem{cohen}
A.~G.~Cohen, D.~B.~Kaplan, F.~Lepeintre and A.~E.~Nelson,
Phys.\ Rev.\ Lett.\  {\bf 78}, 2300 (1997)
[arXiv:hep-ph/9610252] ;
Y.~G.~Kim, P.~Ko and J.~S.~Lee,
Nucl.\ Phys.\ B {\bf 544}, 64 (1999)
[arXiv:hep-ph/9810336].
\bibitem{Cho:2001hx}
G.~C.~Cho, N.~Haba and J.~Hisano,
arXiv:hep-ph/0112163.
\bibitem{belle}
K.~Abe {\it et al.}  [BELLE Collaboration],
BELLE-CONF-0118.
\bibitem{Misiak:1997ei}
M.~Misiak, S.~Pokorski and J.~Rosiek,
Adv.\ Ser.\ Direct.\ High Energy Phys.\  {\bf 15}, 795 (1998)
[arXiv:hep-ph/9703442].
\bibitem{sparticles:limit}
        LEP2 SUSY working group, ALEPH, DELPHI, L3, OPAL experiments,
        {\sf http://lepsusy.web.cern.ch/lepsusy/}.
\bibitem{Chen:2001kn}
C.~H.~Chen and C.~Q.~Geng,
Phys.\ Lett.\ B {\bf 511}, 77 (2001)
[arXiv:hep-ph/0104151].
\bibitem{Arhrib:2001xx}
A.~Arhrib and S.~w.~Baek,
arXiv:hep-ph/0104225.
\end{thebibliography}
\end{document}